\documentclass[10pt,prb,aps,twocolumn,superscriptaddress,showpacs,amsfonts,amsmath,amssymb,showkeys,floatfix]{revtex4-1}
\usepackage{bm}
\usepackage{pstricks}
\usepackage{setspace}
\usepackage{graphicx}% Include figure files
\begin{document}
\title{Numerical results for the Edwards-Anderson spin-glass model at low temperature}
\date{:/fig6\today}
\author{Julio F. Fern\'andez}
\email[E-mail address: ] {jefe@unizar.es}
\affiliation{Departamento de F\'{\i}sica de la Materia Condensada, Universidad de Zaragoza, 50009-Zaragoza, Spain}
\affiliation{Instituto Carlos I de F\'{\i}sica Te\'orica y Computacional,  Universidad de Granada, 18071 Granada, Spain}
\author{Juan J. Alonso}
\affiliation{F\'{\i}sica Aplicada I, Universidad de M\'alaga,
29071-M\'alaga, Spain}
%\altaffiliation{IVIC}

\date{\today}
%\pacs{75.10.Nr,  75.50.Lk, 75.30.Kz, 75.40.Mg}
%\keywords

\begin{abstract}
 We have simulated Edwards-Anderson (EA) as well as Sherrington-Kirkpatrick  systems of
$L^3$  spins. After averaging over large sets of EA system samples of $3\leq L \leq 10$, we obtain accurate numbers for distributions $p(q)$ of the overlap parameter $q$ at very low temperature $T$.
We find $p(0)/T \to 0.233(4)$ as $T\to 0$. This is in contrast with the droplet scenario of spin glasses. 
We also study the number of mismatched links --between replica pairs-- that come with large scale excitations. Contributions from small scale excitations are discarded. We thus obtain for the fractal dimension of outer surfaces of $q\sim 0$ excitations in the EA model $d_{s} \to 2.59(3)$ as $T \to 0$. This is in contrast with $d_s\to 3$ as $T \to 0$ that is predicted by mean field theory for the macroscopic limit. 
\end{abstract}

\maketitle

\section{Introduction}

Whether spin   glasses are complex systems is an important issue. We have discussed this in some detail in Ref. \onlinecite{us},
where we gave numerical evidence for fundamental differences between the  spin-glass phases of the Edwards-Anderson\cite{EA} (EA) and of the Sherrington-Kirkpatrick\cite{SK} (SK) models. In short, we studied  spikes  in probability distributions, $p(q)$, of the overlap parameter, $q$,  that vary widely over different sample systems.  The variation of
a suitably defined average spike width $w$ over the values of the linear system sizes $L$ we studied 
was shown to decrease sharply with $L$ in the SK model. Furthermore, rms deviations $\delta p$ away from $p(q)$  over different system samples (that is, over different realizations of quenched disorder) increase sharply with $L$. Such behavior is consistent with mean field theory, which predicts the replica symmetry breaking (RSB) scenario in which
$w\rightarrow 0$ and $\delta p \rightarrow \infty$ in the macroscopic limit.\cite{seedD,libro}  Our results\cite{us} for the EA model follow a different trend. Rather, $w$ and $\delta p$ become, within errors, independent of $L$  in the zero temperature limit. [The statistics of spikes in overlap distributions in  different system samples, which  have been studied by Yucesoy\cite{yu} et al., also point away from an RSB scenario, though this conclusion is criticized in Ref. \onlinecite{paris}.]
 
Much numerical work on the behavior of the EA model at  low temperature stems from the observations of Moore\cite{bokil} et al., that Monte Carlo simulations had up to then been performed at temperatures that were too close to the critical temperature, and therefore suffered from finite size effects that could be misinterpreted as RSB behavior.  Low temperature data for $p(q)$,
which was the centerpiece of these considerations,
were soon thereafter provided by Katzgraber, Palassini and Young\cite{kpy} (KPY).  A roughly constant value of $p(q\sim 0)$ over a $3\leq L \leq 8$ size range was shown to be consistent with these data. This is as in the RSB,  not  the droplet scenario\cite{droplet,middleton}  of spin glasses. 
 We did not report  data for $p(q)$ in Ref. [\onlinecite{us}], because they were essentially the same as KPY's, and our statistical errors did not decisively improve on them. 
We have since simulated   sample sets which are over an order of magnitude larger than KPY's, and cover a range of system sizes which is slightly larger. We are thus able to report here rather accurate data  for temperatures as low as $0.16T_{sg}$, where $T_{sg}$ is the spin-glass transition temperature.

In the so called trivial-non-trivial (TNT) picture, proposed by Krz\c{a}kala and Martin\cite{halfway} and by Palassini and Young\cite{py} (PY),
$p(q)$ is size independent in the neighborhood of $q=0$, as in the RSB scenario, but the dimensionality $d_s$ of outer surfaces of $q\sim 0$  excitations  is smaller than the dimensionality, $d$, of the space where spins are embedded. Values of $2.57 \lesssim d_s\lesssim 2.62$ have been calculated by PY, KPY, and by J\"org and Katzgraber.\cite{jk}  
However, Contucci et al.\cite{opposite} have obtained $d_s=3$
for the EA model in three dimensions (3D). This would be in accordance with a RSB scenario.  These two conflicting results were obtained by different methods.  Fractions of  mismatched links (FML) between replica pairs are calculated in both methods. 
All $2.57 \lesssim d_s\lesssim 2.62$ values\cite{kpy,py,jk} were obtained (but see Ref. \onlinecite{massive}) from the rms deviation of the FML from its mean value (over time and system samples, as well as over all $q$). On the other hand,
$d_s=3$, was obtained in Ref. \onlinecite{opposite} from the behavior of the mean  FML, $f_{ml}(q)$,  for each observed value of $q$. More specifically,  the $L\rightarrow \infty$ limit of $f_{ml}(q\sim 0)$ was studied in Ref. \onlinecite{opposite} for $T\gtrsim 0.5T_{sg}$.  This limit was argued to be nonzero, which is what one expects of a space fulfilling surface. This conclusion fits with the RSB scenario, and clashes with the ones reached in Refs. \onlinecite{py,kpy,jk}.

Most of  this paper, is devoted to the fractional number of  link mismatches, $\overline{\Delta f_{ml}}^Q$, it \emph{costs} to create an excitation 
with a $-Q<q<Q$ value. For a more precise definition of $\overline{\Delta f_{ml}}^Q$, consider first $\overline{ f_{ml}}^Q$, which is the average FML given that $q$ is in a given $-Q<q<Q$ interval. Subtraction from $\overline{ f_{ml}}^Q$ of the average FML given that $q$ is \emph{not} in the $-Q<q<Q$ interval gives 
$\overline{\Delta f_{ml}}^Q$. Both $\overline{\Delta f_{ml}}^Q$ and 
$\overline{ f_{ml}}^Q$ have  the same zero temperature limit, but we believe $\overline{\Delta f_{ml}}^Q$ is the natural extension to nonzero temperatures of the FML of large-size excitations in the ground state. 
Whereas $\overline {f_{ml}}^Q$ decreases as $T$ decreases, $\overline {\Delta f_{ml}}^Q$ increases. This enables us to bracket very low temperature behavior and confidently make $T\rightarrow 0$ extrapolations.  These notions stand out clearly  in the frustrated box (FB) model which we define below.

The plan of this paper is as follows. 
In Sec. \ref{method}, we define the models, the spin-overlap  and link-overlap parameters, and the simulation procedure. 
In Sec. \ref{averagep}, we report accurate data for $p(q)$ for  EA and SK systems at very low temperature.
These data show  that, in the $3\leq L\leq 10$ range, $p(q\sim 0)$ is independent of $L$ at  very low temperatures. The conclusion KPY\cite{kpy} reached, that the EA model exhibits a clear trend away from the droplet scenario, is thus strengthened.
In Sec. \ref{fb} we examine the large-scale behavior of the FML in the FB model.  This simple model helps to highlight the pitfalls that should  be avoided in the  interpretation  of a nonzero macroscopic limit of FML.  
In Sec. \ref{deltafml}, we assign a (average) mismatching-link cost, $\overline{\Delta f_{ml}}^Q$, to an excitation with a $-Q<q<Q$ value. 
Numerical results for $\overline{\Delta f_{ml}}^Q$, which imply a fractal dimension of $2.59(3)$ for the surface associated to $\overline{\Delta f_{ml}}^Q$, are also given in Sec. \ref{deltafml}.  Such a value of $d_s$, smaller than the dimensionality 3D of the space where spins are embedded, is in contradiction with mean field theory predictions, but
is as envisioned in the TNT scenario\cite{halfway,py} of the EA model. 
We summarize our conclusions in Sec. \ref{conclusions}.

\section{Models, definitions and procedure}
\label{method}

 In all models we study, an Ising  spin sits on each one of the $N\equiv L\times L\times L$ sites of a simple cubic lattice in three dimensions (3D). We use periodic boundary conditions throughout.
In  the SK and EA models the interaction energy between a pair of spins at sites $i$ and $j$ is given by $J_{ij}\sigma_i\sigma_j$. We let $J_{ij}=\pm 1/\sqrt{N}$ randomly, without bias, for all $ij$ site pairs in the SK model. For the EA model,  $J_{ij}=0$ unless $ij$ are nearest-neighbor pairs, and we draw
each nearest-neighbor bond $J_{ij}$  independently from unbiased Gaussian distributions of unit variance. 

We let all temperatures be given in units of $1/k_B$, where $k_B$ is Boltzmann's constant.
Thus, the transition temperature $T_{sg}$ between the paramagnetic and SG phase of the SK model is $T_{sg}=1$. \cite{libro,SK}
For the EA model $T_{sg}\simeq 0.95$. \cite{TcEA}

We let  $\sigma_i^{(1)}$ stand for a spin at site $i$ of replica  $1$ of a given system,  and similarly, $\sigma_i^{(2)}$  for an identical replica, replica $(2)$, of the same system. As usual, we define 
\begin{equation}
 q\equiv N^{-1}\Sigma_i q_i  \;\text{and}  \;  q_i=\sigma_i^{(1)}  \sigma_i^{(2)},
 \label{q}
\end{equation}
that is, $q$ is the average (over all sites) spin alignment between the states replicas $1$ and $2$ are in. 

As in Refs. \onlinecite{halfway,py}, we define the link-overlap,
\begin{equation}
 q_l\equiv (N_l)^{-1}\Sigma_{\langle  ij \rangle }\sigma_i^{(1)}\sigma_j^{(1)}   \sigma_i^{(2)}\sigma_j^{(2)},
 \label{f}
\end{equation} 
where $N_l$ is the total number of links, and the sum is over all $i,j$ nearest neighbor pairs (of which there are $3N$ in the nearest neighbor EA model in 3D). The FML between replicas $1$ and $2$ is given by $f_{ml} = (1-q_l)/2$. 
In addition, as in Ref. \onlinecite{opposite}, we define $f_{{\cal J}ml}(q)$ as    
the time average of the FML  (for a sample with a given set ${\cal J}$ of bonds) which is observed over all   time intervals while the value of the spin-overlap is $q$.
Unfortunately,  the dimensionality $d_s$ of large scale excitations does not follow straightforwardly from the behavior of $f_{{\cal J}ml}$. 
This is because  smaller scale excitations contribute to $f_{{\cal J}ml}(q)$. More on this can be found in  Sec. \ref{df}.

Let ${\cal F}(q)$ be any $q$ dependent function, such as $p(q)$ or $f_{ml}(q)$. We define
\begin{equation}
\overline{{\cal F}}^Q \equiv (2Q)^{-1}\int_{-Q}^Q dq{\cal F}(q).
\end{equation}
The advantage of working with $\overline{{\cal F}}^Q$ is that statistical errors for it are smaller than for ${\cal F}(q)$.
Accordingly, most results below are given for $\overline {p}^Q$ and $\overline{f_{ml}}^Q$ rather than for $p(q)$ and $f_{ml}(q)$. How statistical errors on $\overline {p}^Q$ depend on $Q$  is worked out in Appendix \ref{confidence}.

In addition, we let $A_{\cal J}$ stand for the value of some observable $A$ on a sample defined by the set ${\cal J}$ of bonds, and we let $\langle  A_{\cal J} \rangle_{\cal J}$ stand for the average over samples of $A_{\cal J}$.
Thus, $p(q)\equiv \langle p_{\cal J}(q) \rangle_{\cal J}$ and
 $[\delta p(q)]^2  \equiv \langle [p_{\cal J}(q)-p(q)]^2 \rangle_{\cal J}$.

\begin{table}\footnotesize
\begin{center}
%\begin{ruledtabular}
\vspace{0.5 cm}
\begin{tabular}{|c| r  r r  |r r rr r | cr|}
\hline
 & & SK&  & &  &  EA     & & & FB & \\
\hline
L & 4 & 6 & 8  & 4& 5 &6 &8 & 10 & 4-12& \\
\hline
$\tau_s$ & $10^5$ & $ 10^5$ & $10^5$ &$10^4$ &$10^4$ & $10^5$ &$10^6$ & $10^7$  &$10^6$& \\
$N_s$ & $10^5$    & $10^5$ & $10^5$ & $10^5$ & $10^5$ & $10^5$ &$10^5$  & $2\times 10^4$ & $1$ &\\
$\alpha$ & $0.7$  & $0.5$ &  $0.4$ & $0.7$ &  $0.6$ & $0.5$ & $0.5$  & $0.45$ &$ >0.3$ & \\
\hline
\end{tabular}
%\end{ruledtabular}
\caption{Number of samples,  $N_s$, and the number of MC sweeps  $\tau_s$   taken  for equilibration as well as  for  subsequent averaging.  The average over samples of the swap success rate, $\alpha$, between EA and SK systems at their two lowest temperatures, and between systems in the critical region of the FB system, are also given. (Other swap rates are larger.)}
\end{center}
\end{table}

We make use of the parallel tempered MC method.\cite{tmc,tmc3} Details on how we apply it to the EA and SK model are as specified in Ref. [\onlinecite{us}].
However, some details differ.
For all  sizes of the EA model, temperatures are spaced here by $0.04$ ($0.08$) in the  $0.16 \leq T\leq 0.48$ ($0.48 \leq T \leq 1.6$) range.
The rationale for this, as well as checks we perform in order to make sure equilibrium is reached, can be found in Appendix \ref{swaprates}.
Temperatures of all SK systems  were evenly spaced by $\Delta T=0.04$ in the whole $0.12 \leq T \leq 1.6$ range. For the FB model, $\Delta T=0.02$ ($\Delta T=0.01$) for all $4\leq L \leq 8$ ($L=12$). 
Values for average swap success rates, $\alpha$, between pairs of EA and SK systems at the lowest two temperatures are given in Table I. Larger values of $\alpha$ are observed for higher temperatures. In the FB model, the smallest value of $\alpha$, which we give in Table 1, is observed in the critical region. 
 
The number, $N_s$, of sample systems we average over, is, as specified in Table I, much larger here than in Ref. [\onlinecite{us}]. We have tried not to make $N_s$ smaller with increasing $L$. This is because, as we show in Appendix \ref{confidence}, statistical errors are independent  of $L$, because of non-self-averaging. (For $L=10$, we could only do $20\;000$ samples. That took some $50 $ years worth of computer time.)

\begin{figure}[!t]
 \begin{center}
\includegraphics*[width=80mm]{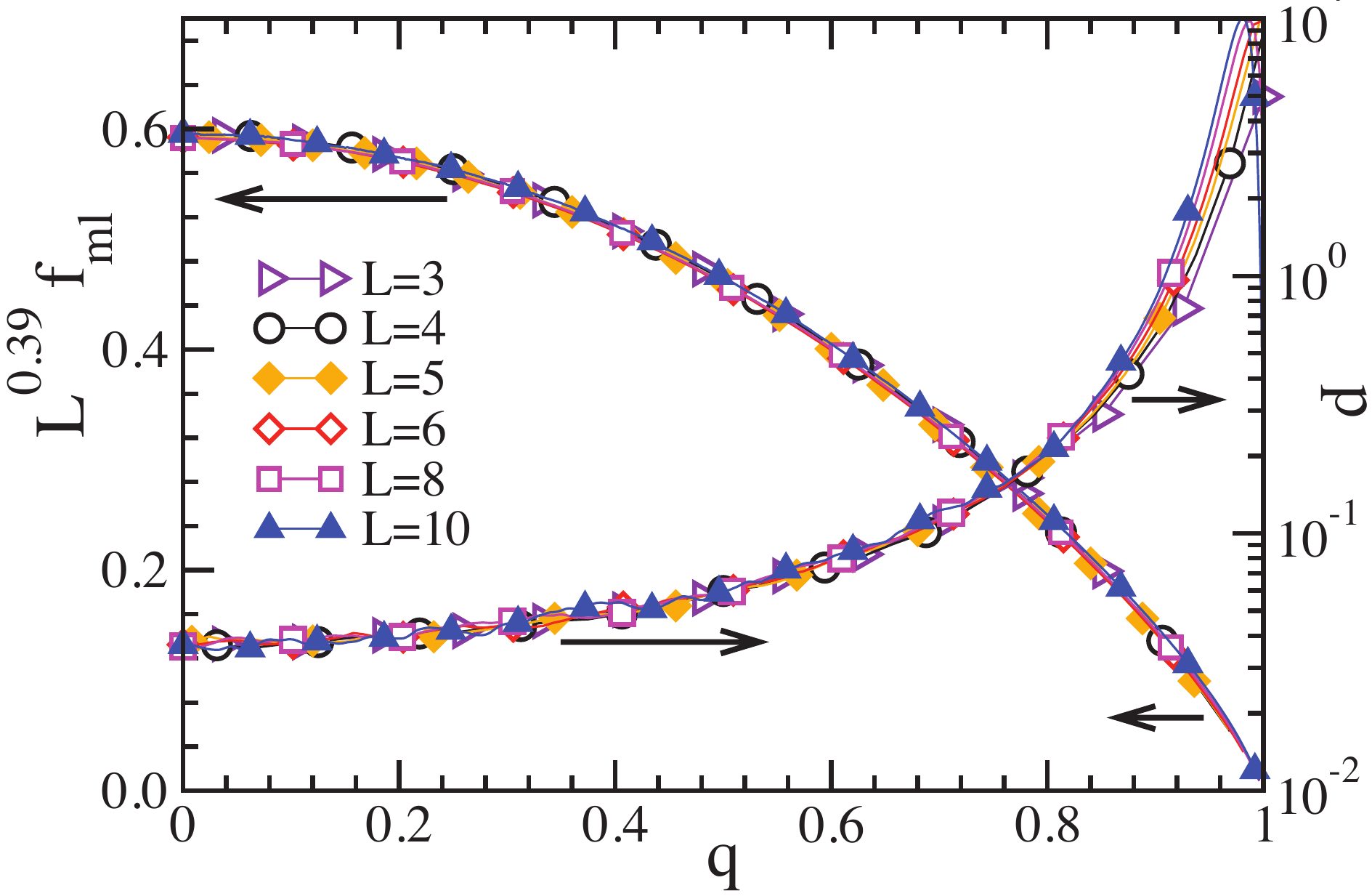}
\caption{(Color online) Plots of $L^{0.39}f_{ml}$  and of $p$ vs $q$,  for EA systems of $L^3$ spins, for the values of $L$ shown, at $T=0.16$.
For clarity, only a fraction of all data points are shown, but the lines shown go through every one of them.
The normalization condition $\int_{-1}^1dq\;p(q)=1$ is satisfied.}
\label{links}
\end{center}
\end{figure}

\section{Average $q$-distributions at low temperatures}
\label{averagep}

\begin{figure}[!t]
 \begin{center}
\includegraphics*[width=75mm]{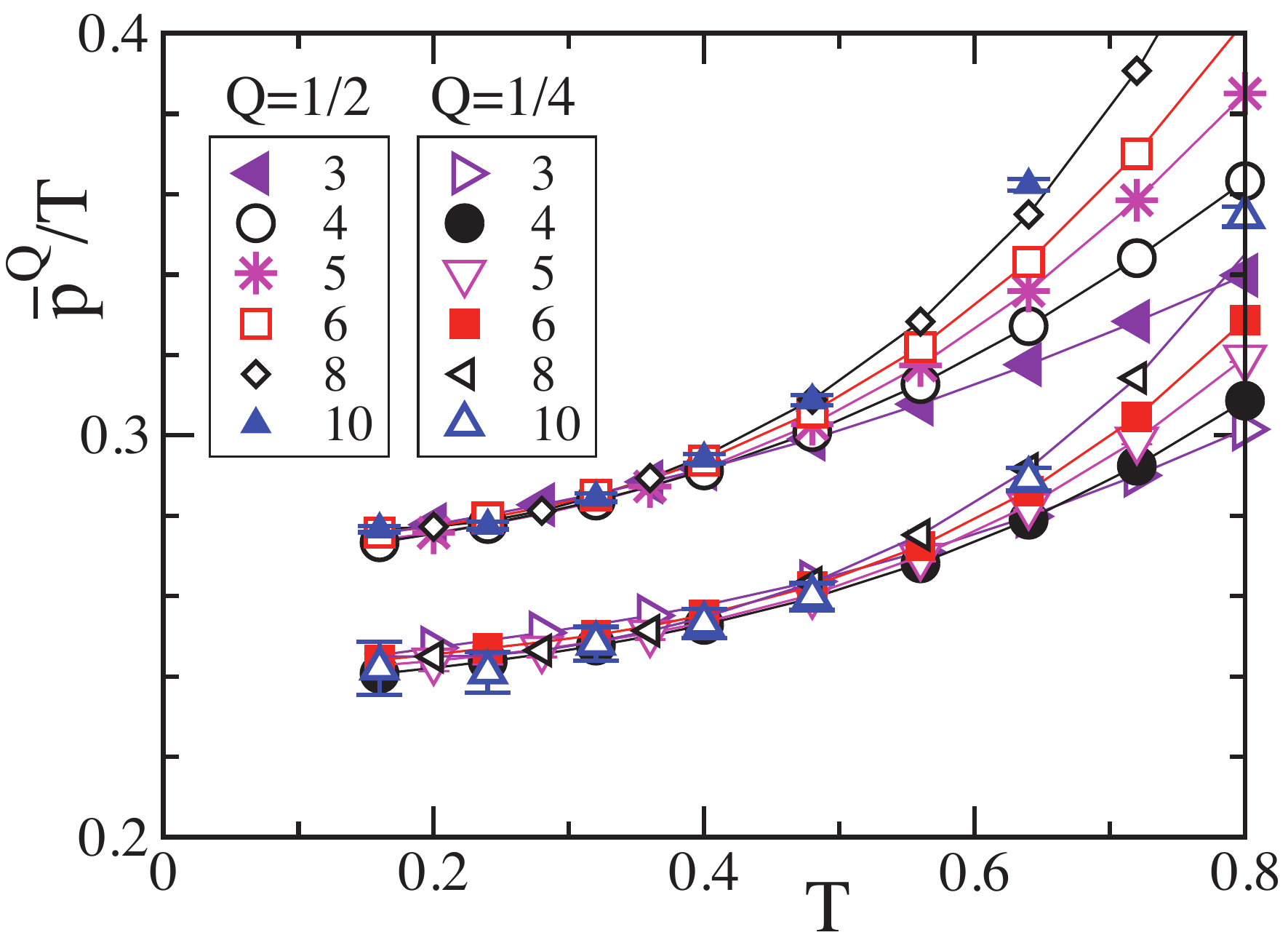}
\caption{(Color online) Plots of $\overline{p}^Q$ vs $T$ for EA systems of $L\times L\times L$ spins, for the values of $L$ shown, and $Q=1/4$ and $1/2$. Icons for all $L\leq 8$ cover their error bars.  For better visibility of data points, not all of them are shown, but lines go through every one of them.}
\label{pvsT}
\end{center}
\end{figure}

Plots of $p(q)$ vs $q$ are shown in Fig. \ref{links} for  EA systems of various sizes at $T=0.16$.  Plots of
$\overline{p}^Q/T$  vs $T$ are   shown in Fig. \ref{pvsT}  for $Q=1/4$ and $1/2$.  Error sizes are clearly smaller for the larger value of $Q$. Simulation details, such as  sample numbers and running times, are given in Table I.

\begin{figure}[!t]
 \begin{center}
\includegraphics*[width=75mm]{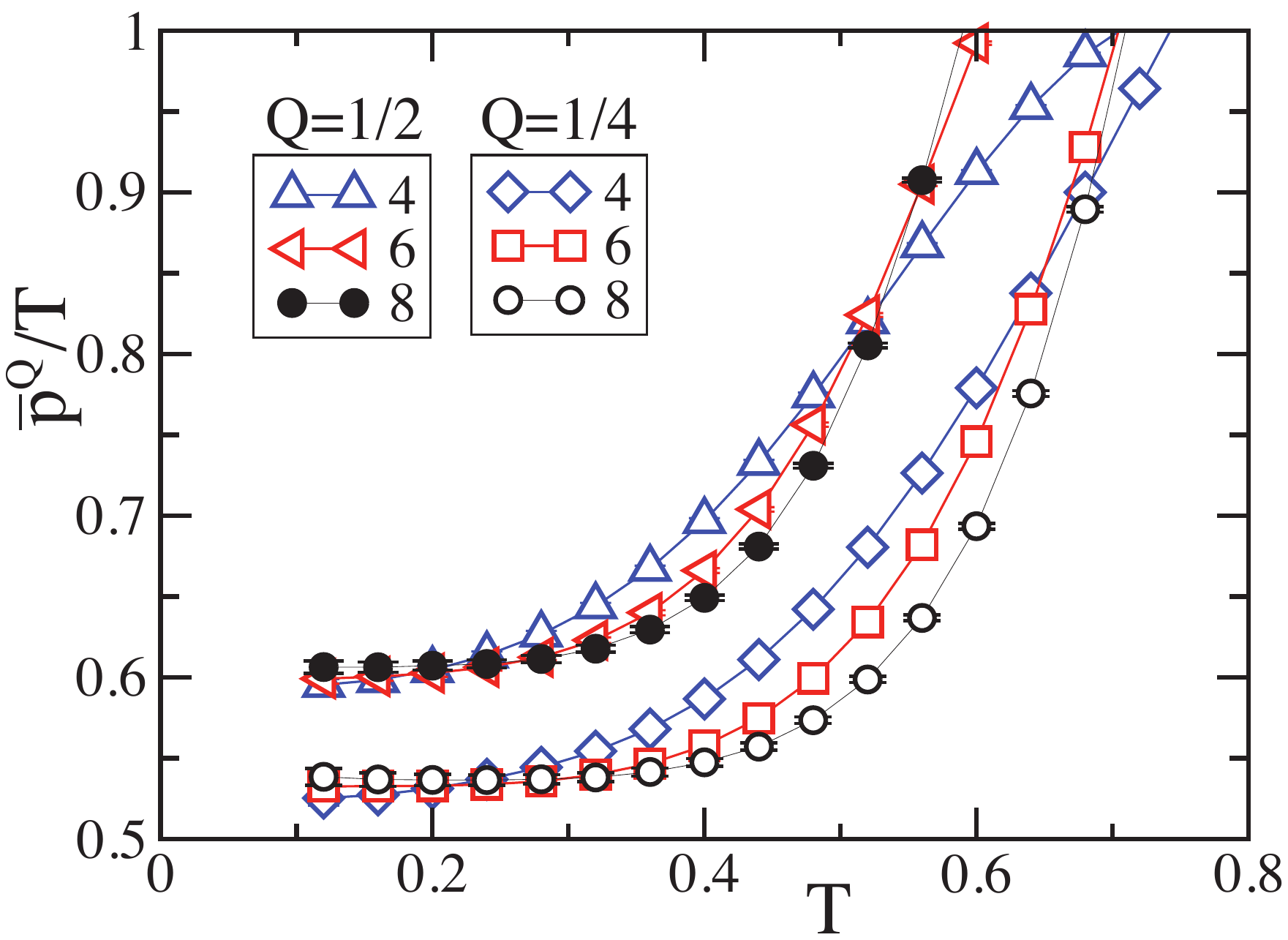}
\caption{(Color online) (a) Plots of $\overline{p}^Q$ vs $T$ for SK systems of $L\times L\times L$ spins, for the values of $L$ shown, and $Q=1/4$ and $1/2$. Icons cover all error bars.}
\label{SK}
\end{center}
\end{figure}

For comparison, plots of  $\overline{p}^Q/T$  vs $T$  for SK systems of various sizes are shown in Fig. \ref{SK} for $Q=1/2$ and $1/4$.
We note   that  $\overline {p}^Q \sim T$ if $T\lesssim 0.3T$, independently of $L$, following  mean field predictions,\cite{libro,vani} for the SK model.

\begin{figure}[!t]
 \begin{center}
\includegraphics*[width=75mm]{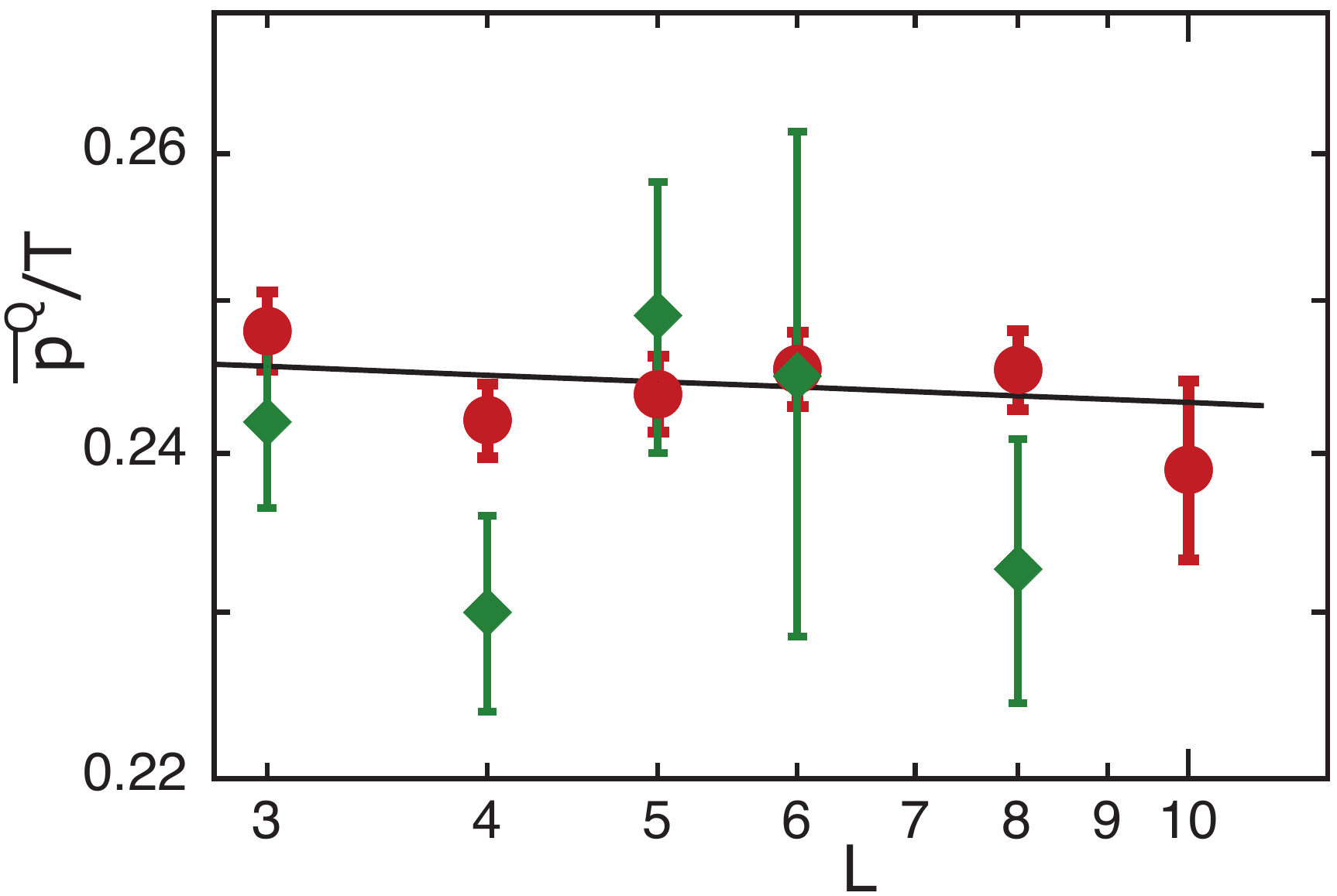}
\caption{(Color online) Log-log plots of $\overline{p}^Q$ vs $L$  for $Q=1/4$ of EA systems at $T=0.2$.
All $\bullet$ stand for our own MC data and the straight line for the best  $\overline{p}^Q \sim L^{-\theta }$  fit (given by $\theta \simeq 0.008$) to the data. The best fit gives $\chi^2\simeq 3.7$, and  $\chi^2\simeq 10$ ($\chi^2\simeq 8$) follows from letting $\theta = 0.04$ ($-0.02$). All $\blacklozenge$  stand for old data points from KPY\cite{kpy} for the same temperature and a slightly lower $Q=0.2$ value. [We have divided by $2$ data values from KPY, because their reported values of $p(q)$  fulfill $\int_0^1dq\;p(q)=1$, instead of  $\int_{-1}^1dq\;p(q)=1$.]
 }
\label{inset}
\end{center}
\end{figure}

For a more accurate picture of  how $\overline{p}^Q$ varies with $L$ in the EA model, we show log-log plots of  $\overline{p}^Q/T$ vs $L$, for $Q=1/4$ and $T=0.2$ in  Fig. \ref{inset}. For comparison, we also show data points from the KPY paper for the same temperature and $Q=0.2$.\cite{Q}

The best fit of  $\overline{p}^Q\propto L^{-\theta}$ to the data points shown in Fig. \ref{inset} gives $\theta\simeq 0.008$. Fits following from letting $\theta =0.04$ and $-0.02$ give $\chi^2$ parameters that are over twice as large as the one for $\theta = 0.008$. (See the figure legend for further details.) For higher temperatures, up to $T\simeq 0.4$, as well as for $Q=1/2$ and all $ T \lesssim 0.5$,   all error bars are smaller than the ones shown in Fig. \ref{inset}, and
 all best fits of  $\overline{p}^Q\propto L^{-\theta}$ to the data give $\mid \theta \mid <0.01$.  Thus, future generation of more accurate data that would give $\theta > 0.04$ for the $3\leq L \leq 10$ range is rather unlikely.

Finally, $Q\rightarrow 0$ and $T\rightarrow 0$ extrapolations, give $p(0)/T \rightarrow 0.51(3)$ for the SK model. Similarly,  $p(0)/T \rightarrow 0.233(4)$ as $T\rightarrow 0$ follows from the plots shown in Fig. \ref{pvsT}  for the EA model.

\section{Number of link mismatches which come with  large scale excitations} 
\label{df}

In this section we give a definition of the fraction of mismatches, $\overline {\Delta f_{ml}}^Q$, it costs  to create an excitation in the $-Q<q<Q$ range, where $0<Q<1$.
The simplicity of the FB model is helpful in this respect. We introduce this nonrandom frustrated 
model in Sec. \ref{fb}. The definition of $\overline {\Delta f_{ml}}^Q$ as well as the results we obtain for the EA (and SK) model are given in Sec. \ref{deltafml}.

\subsection{The frustrated-box model}
\label{fb}

We define here a nearest-neighbor Ising model in which most bonds are ferromagnetic. For reasons given below, we term it the \emph{frustrated-box} (FB) model.  Consider plane $\mathfrak{ P}_1$, perpendicular to the $x$ axis, at $x=1/2$, 
which cuts all bonds between $x=0,y,z$, and $1,y,z$ sites. Similarly, $\mathfrak{ P}_2$ at $x=L/2+1/2$,  cutting all bonds between $x=L/2,y,z$ and $L/2+1,y,z$. These planes divide the system into two equal portions. In this model,  only nearest-neighbor spins interact.
All bonds, except the ones that cut across $\mathfrak{ P}_1$ and $\mathfrak{ P}_2$, are of strength $1$, that is, ferromagnetic.  Half the bonds that cut across both $\mathfrak{ P}_1$ and $\mathfrak{ P}_2$ are of strength $-1$,  that is, antiferromagnetic, and the rest are of strength $1$. More precisely, all $\pm 1$  bonds that cut across both $\mathfrak{ P}_1$ and $\mathfrak{ P}_2$ are distributed on a checkerboard pattern. We apply periodic boundary conditions.

In the ground state, all spins within the box (that is, between planes $\mathfrak{ P}_1$ and $\mathfrak{ P}_2$) are parallel, and so are all spins outside the box. These two spin subsystems can point in the same or opposite directions.
Thus, the ground state is (because of invariance under all-spin reversal) four-fold degenerate.
The box defined by $\mathfrak{ P}_1$ and $\mathfrak{ P}_2$ and the system's boundary is the 3D analog of Toulouse's two-dimensional frustrated plaquettes.\cite{toulo}
Hence, the ``frustrated-box" label.

\begin{figure}[!t]
 \begin{center}
\includegraphics*[width=75mm]{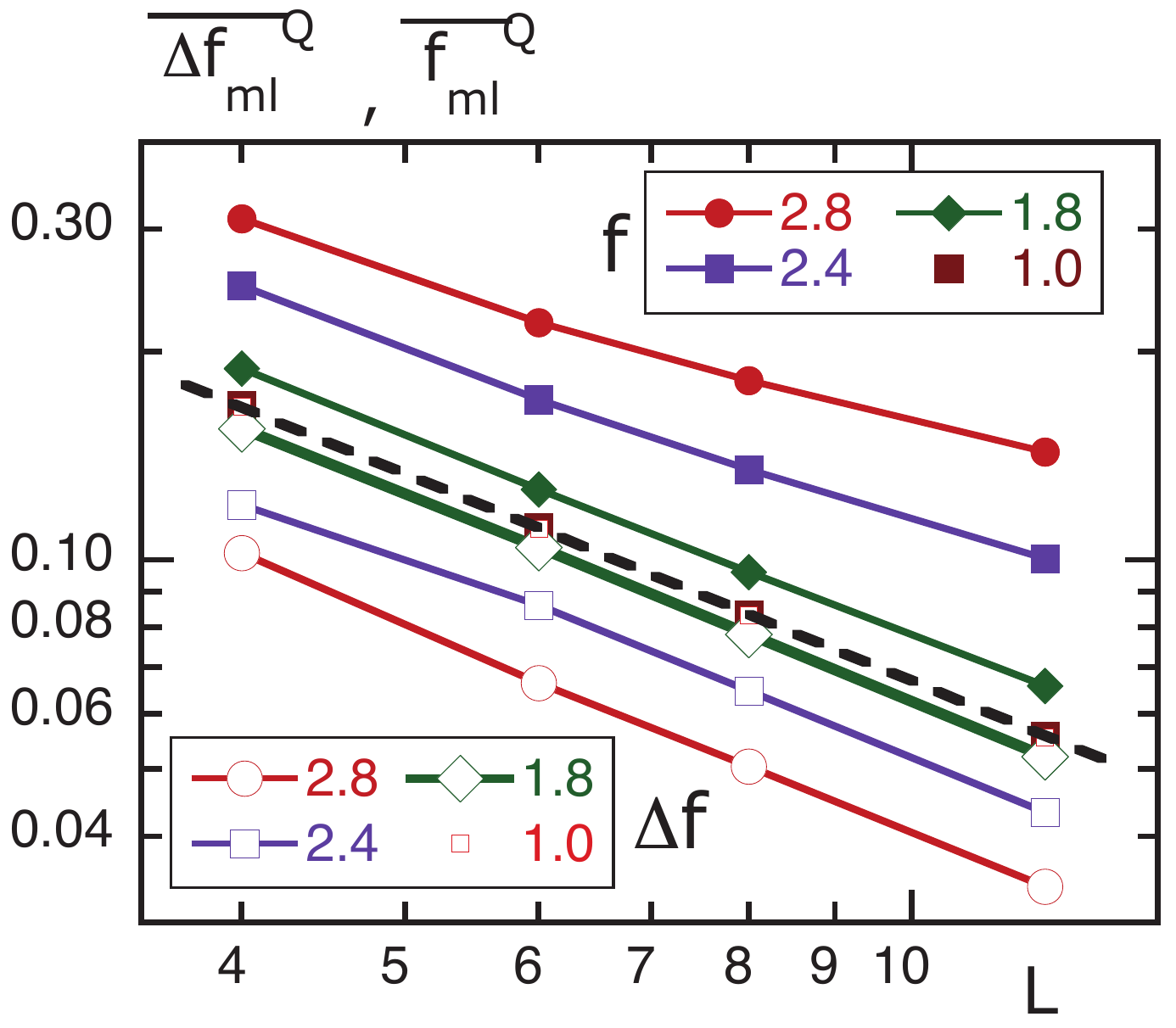}
\caption{(Color online) Plots of $\overline{\Delta f_{ml}}^Q$ and of $ \overline{f_{ml}}^Q$, for $Q=1/2$, vs $L$ for FB systems at the $T$ values shown. The dotted line is for $2/3L$.
In the parallel tempered MC setup, $L=8$ FB systems were placed in thermal contact with heat baths at equally spaced, by  $\Delta T$, 
in the $6.5\geq T \geq 0.8$ temperature range. For all $L\leq 8$, $\Delta T=0.02$, but $\Delta T=0.01$ for $L=12$. }
\label{2plfvsL}
\end{center}
\end{figure}

The number of \emph{broken} bonds in all ground states of the FB model is $L^2$, but the number of bond \emph{mismatches}   between two replicas is (in ground states) either $0$ or $2L^2$. Thus, ${f_{ml}}=1/3L$ but $f_{ml}(q=0)=2/3L$. Plots of  $\overline{f_{ml}}^Q$ vs $L$ are shown in Fig. \ref{2plfvsL} for $Q=1/2$  in  FB systems at
various temperatures. Note $\overline{f_{ml}}^Q=2/3L$  at $T=1$, as expected for $T\ll T_{c}\simeq 4.5$. 

Curves for $T=2.4$  and $2.8$ in Fig. \ref{2plfvsL} clearly hint at a nonzero asymptotic value of $\overline{f_{ml}}^Q$. 
The right interpretation of this result comes easily for the FB model. Obviously, it isn't that $d_s\rightarrow 3$ as $L\rightarrow \infty$. Rather, bulk contributions to $\overline{f_{ml}}^Q$, compete with contributions (amounting to $2/3L$) from the outer surfaces enveloping $q\sim 0$ excitations when $L$ is sufficiently large. Consider, for instance, $T\lesssim 2.5$. A $\exp (-12/T)$ fraction  of all spins point in the ``wrong'' direction then in large FB systems. This is the reason why $\overline{f_{ml}}^Q$ must cross over to a size-independent value  [at $L\sim (1/18)\exp (12/T)$ in the FB model]. Below, we subtract from $\overline{f_{ml}}^Q$ unwanted contributions.  

\begin{figure}[!b]
%\begin{center}
\includegraphics*[width=70mm]{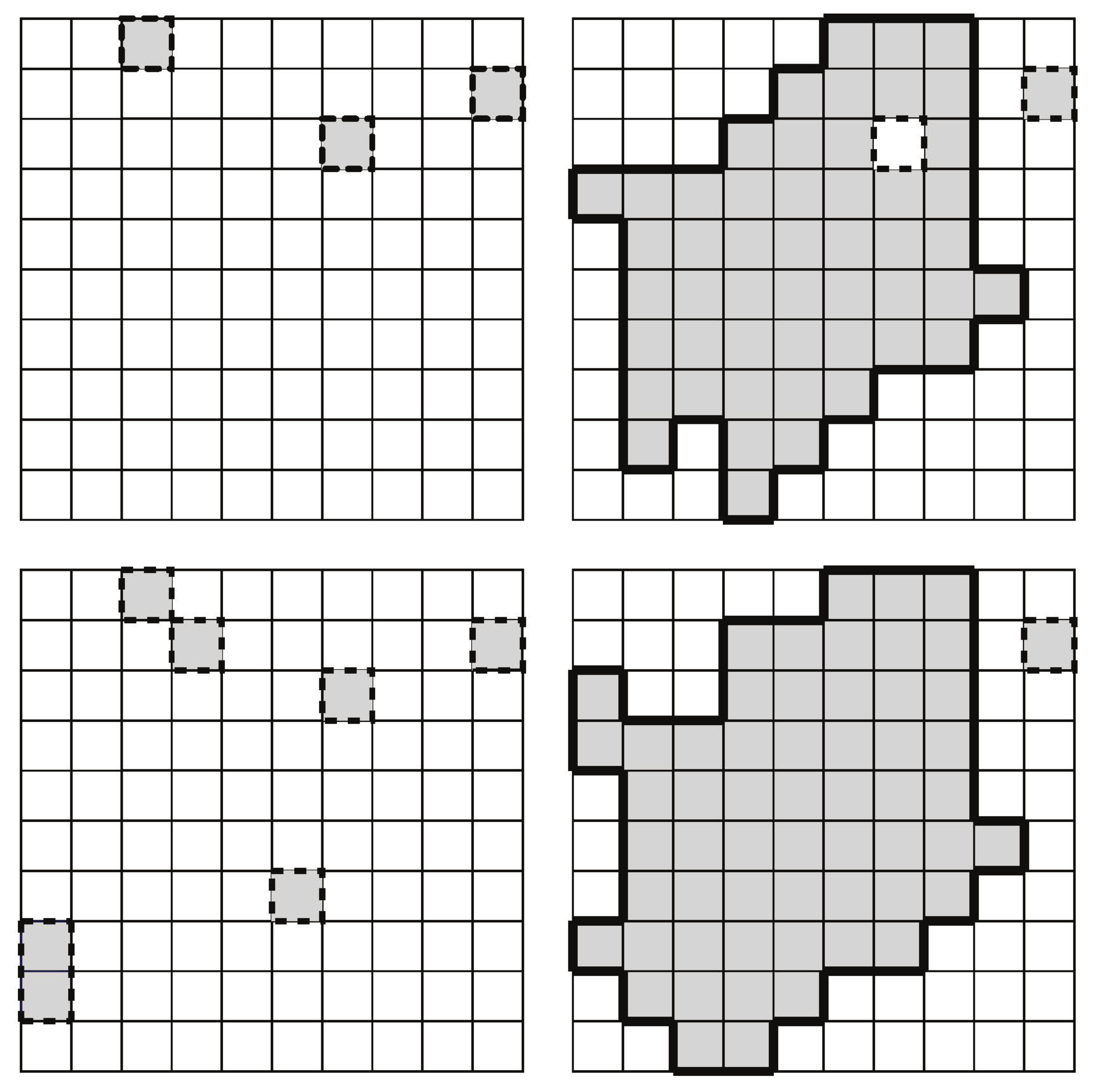}
\caption{Maps showing grey (white) squares for sites where $q_i=-1$ ($q_i=+1$) on a horizontal cut of  
a $10\times 10\times 10$ spin system. Outer surfaces of large scale excitations are shown with thick black lines.
Broken lines stand for surfaces of small scale excitations.
Different panels show maps observed at different times (consecutive times are at least $10^6$ MC sweeps apart) from a simulation of an EA sample system at $T=0.36$. The spin-overlap parameter, taken over the whole system, had values $q\simeq 0.98, 0.88, -0.06, -0.06$ for the top left, bottom left, top right and bottom right panels, respectively, at the times these images were recorded. }
\label{2D}
%\end{center}
\end{figure}

\subsection{Number of link mismatches which come with  large scale excitations in the EA model} 
\label{deltafml}

\begin{figure}[!t]
%\begin{center}
\includegraphics*[width=70mm]{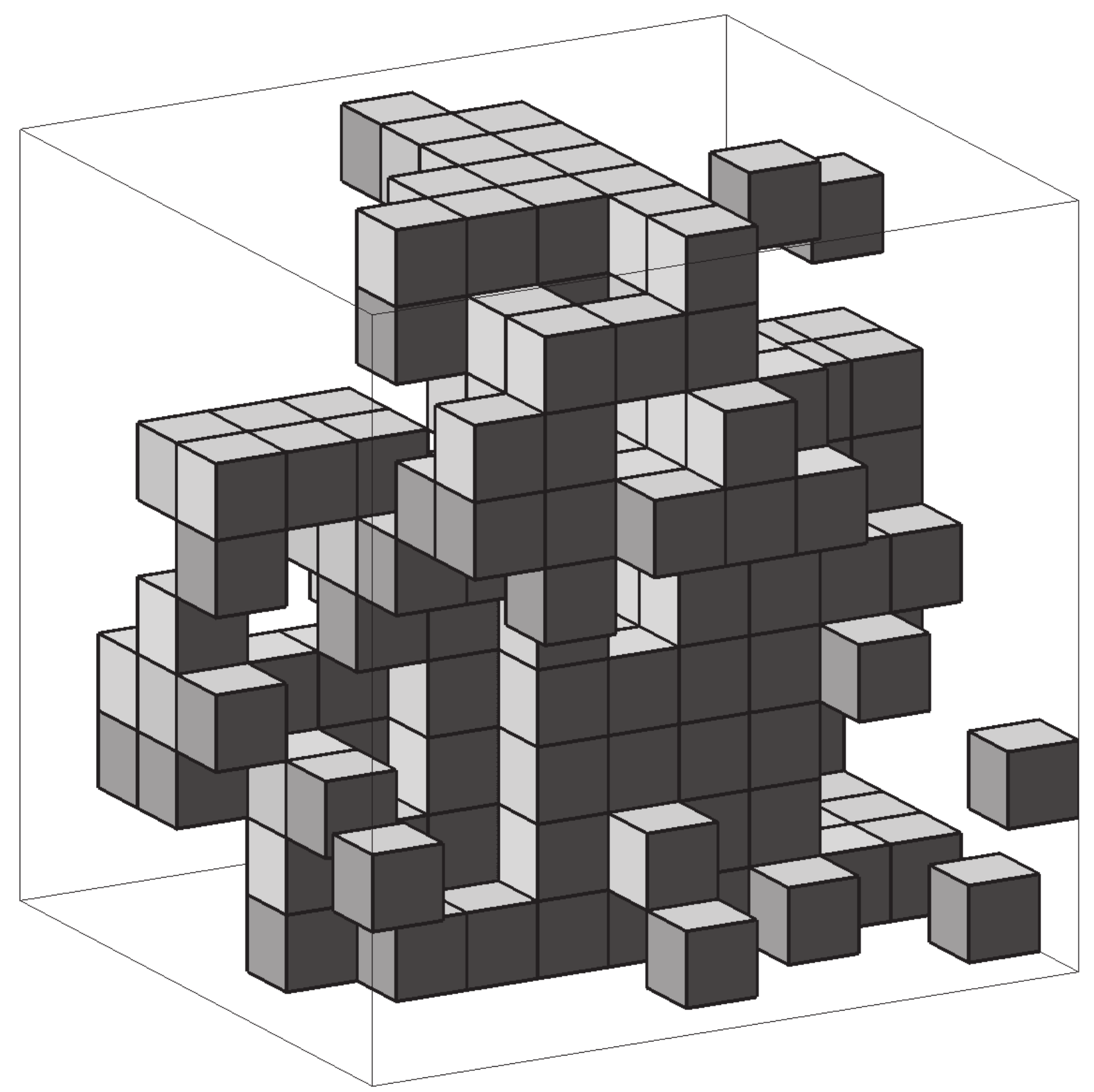}
\caption{Instantaneous 3D image showing outer surfaces on several small- and one large-scale excitation  in an EA system sample at $T=0.16$. Flat surfaces on the framing cubic box are not excitation boundaries and should be disregarded as such. They arise from our inability to give an appropriate pictorial representation of periodic boundary conditions.
At the time this image was recorded, $q\simeq 0.495$.}
\label{3Da}
%\end{center}
\end{figure}

Let us first examine a simple picture of large and small scale excitations.
In Fig. \ref{2D}, the same cross section of an EA sample system of $10^3$  spins at $T=0.36$ is shown at four different times (consecutive times are at least $10^6$ MC sweeps apart) of a single MC simulation. 
For a 3D picture of the outer surface of a large scale excitation in an EA system at $T=0.16$ see Fig. \ref{3Da}.

The general idea is to determine the  area of the outer surface of large scale excitations, such as the ones on both right-hand panels of Fig. \ref{2D}. We intend to do this by subtracting the total surface area of small scale excitations from the total (from small and large excitations) surface area.  

For each system sample, we first obtain $f_{{\cal J}ml}(q)$ for each $q$  by  adding $1/2(1-q_l)$ whenever $q$ is observed in a given MC run, and we finally divide the result by the number of times $q$ has been observed. 
Now, the average surface area over all excitations observed in a given sample whenever a value of $q$ is in the
$-Q<q<Q$ range is given by,
\begin{equation}
\overline {f}^{Q}_{{\cal J}ml} \equiv (1/u_{\cal J})\int_{-Q}^Qdq \;f_{{\cal J}ml}(q)p_{\cal J}(q) ,  
\label{f3}
\end{equation}
where $u_{\cal J}=\int_{-Q}^Qdq \;p_{\cal J}(q)$, 
and by,
\begin{equation}
\overline {f}^{{\bf\not}{Q}}_{{\cal J}ml} \equiv (1/v_{\cal J}) \int_{\mid q \mid > Q}dq \;f_{{\cal J}ml}(q)p_{\cal J}(q)  ,
 \label{fneg}
 \end{equation}
where $ v_{\cal J}=\int_{\mid q \mid > Q}dq \;p_{\cal J}(q)$, whenever $q$ is \emph{outside} the $-Q<q<Q$ range.

Finally,  for the average FML it \emph{costs} to create an excitation in the $-Q<q<Q$ range, we calculate,
\begin{equation}
\overline {\Delta f_{ml}}^Q \equiv \langle \overline { f_{{\cal J}ml}}^Q  -  \overline { f_{{\cal J}ml}}^{\not{Q}}\rangle_{\cal J}.
\label{abc}
\end{equation} 
We have calculated $\langle \ldots \rangle_{\cal J}$ in the above equation by each of the following two procedures: (1) giving equal weight 
to all system samples for which $u_{\cal J}>0.1$, and (2) giving    each sample a weight proportional to $\int_{-Q}^Qdq \;p_{\cal J}(q)$. 
Within statistical errors, we have obtained the same results from these two procedures.

Thus, $N_l\overline {\Delta f_{ml}}^Q$ is a reasonable definition of the outer surface area $\mathfrak {S}$  of an excitation with a  $-Q<q<Q$ value. This definition excludes contributions from small scale excitations.

\begin{figure}[!t]
 \begin{center}
\includegraphics*[width=80mm]{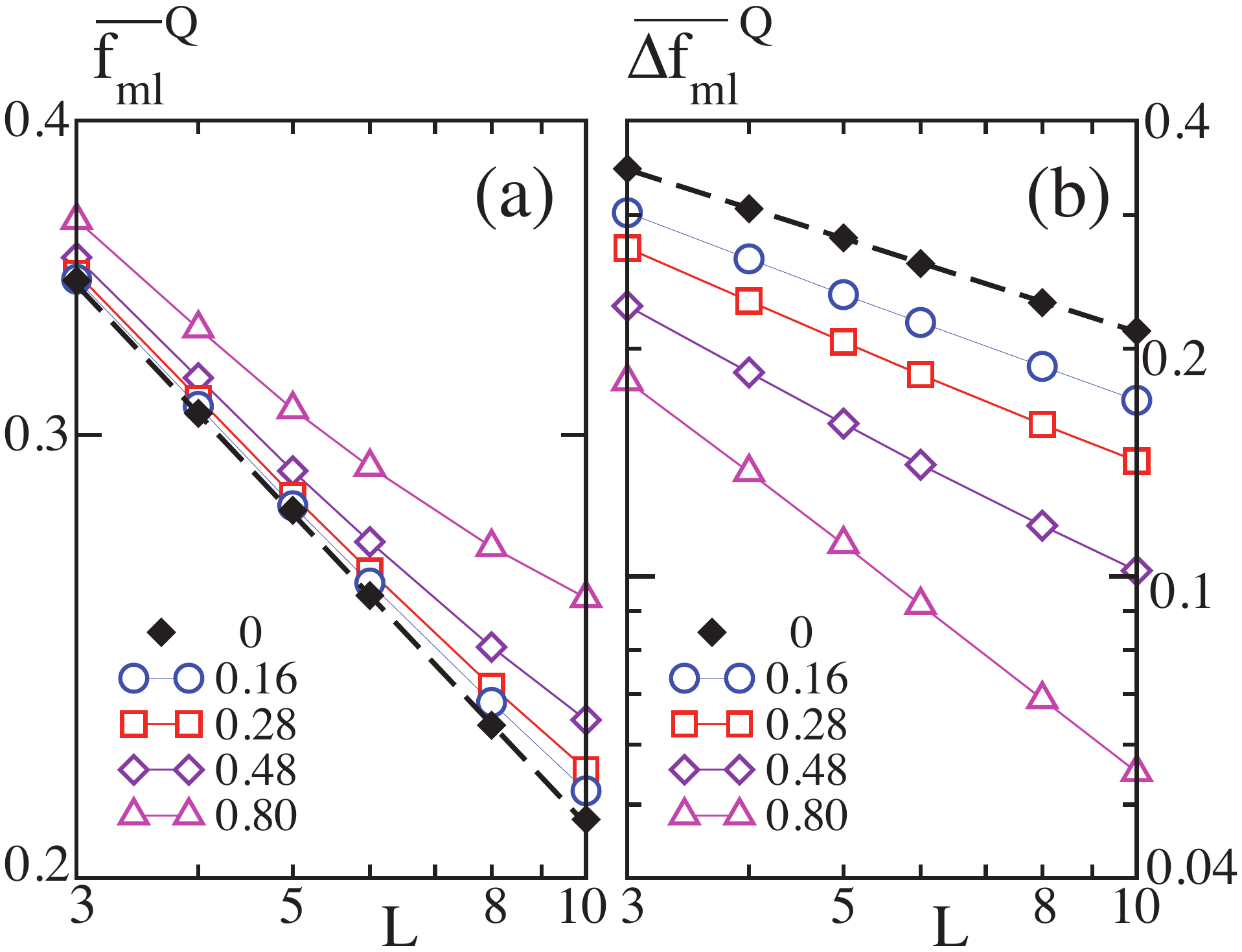}
\caption{(Color online)  (a) Plots of $\overline{ f_{ml}}^Q$  vs $L$,  where $Q=1/2$, for  EA systems of $L^3$ spins, for the values of $T$ shown. The data points ($\blacklozenge$) shown for $T=0$ come from extrapolations such as the ones shown in Fig. \ref{bT}. The slope of the dashed line going through the $T=0$ data points is $0.41$. (b) Same as in (a) but for $\overline{\Delta f_{ml}}^Q$. The $\blacklozenge$ 
 data points are the same as in (a). }
\label{linksLm}
\end{center}
\end{figure}

We can  first check in Fig. \ref{2plfvsL} for the general behavior of $\overline {\Delta f_{ml}}^Q$ in the FB model.  Data points approximately fall on straight lines. Furthermore,  all are well fitted by $\overline {\Delta f_{ml}}^Q \sim 1/L$, thus giving the desired value, $d_s=2$, for the dimension of planes $\mathfrak{P}_1$ and $\mathfrak{P}_2$, not only as $T\rightarrow 0$ but for  nonzero temperatures as well. 

\begin{figure}[!b]
 \begin{center}
\includegraphics*[width=80mm]{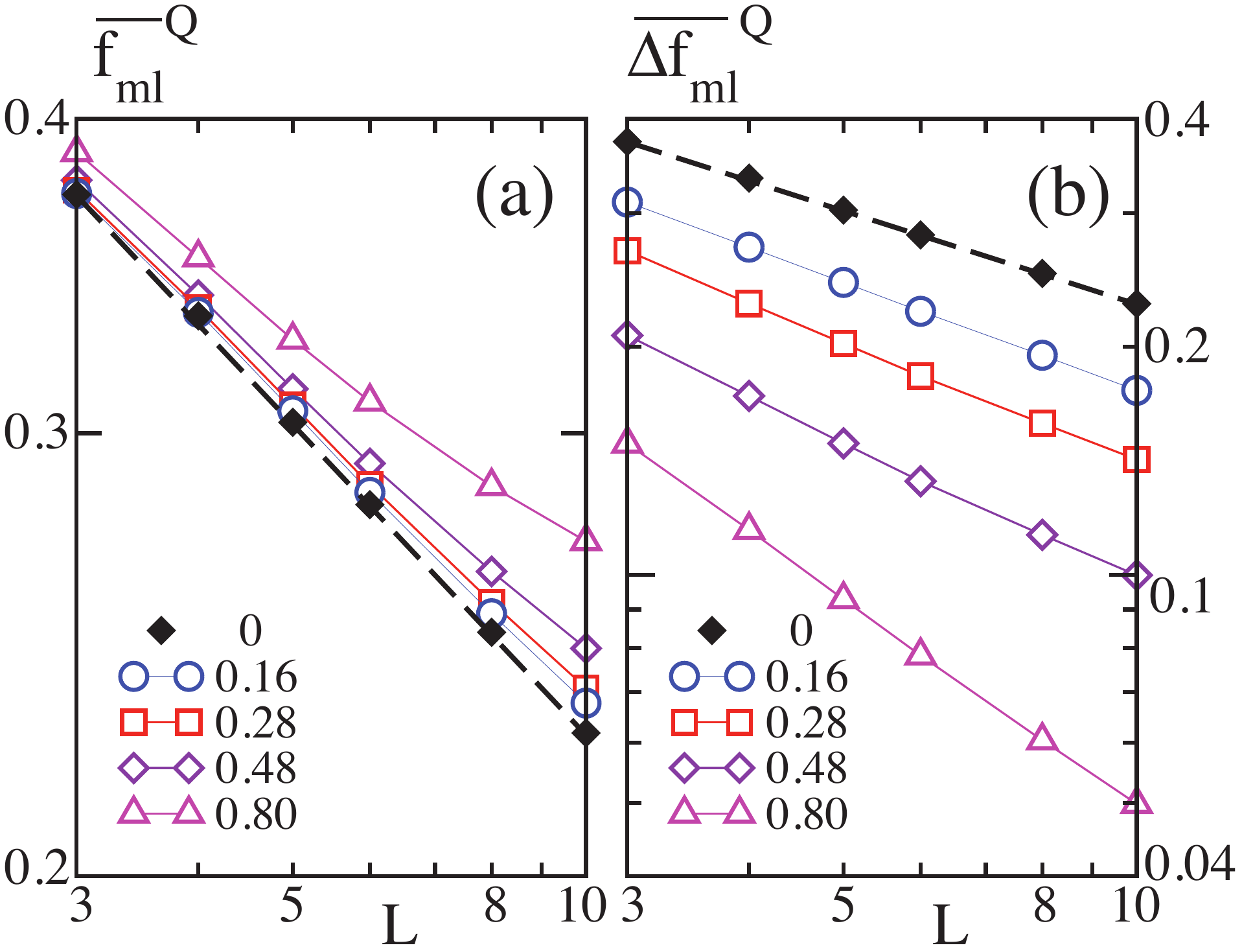}
\caption{(Color online)  (a) Plots of $\overline{ f_{ml}}^Q$  vs $L$,  where $Q=1/4$, for  EA systems of $L^3$ spins, for the values of $T$ shown. The data points ($\blacklozenge$) shown for $T=0$ come from extrapolations such as the ones shown in Fig. \ref{bT}. The slope of the dashed line going through the $T=0$ data points is $0.41$. (b) Same as in (a) but for $\overline{\Delta f_{ml}}^Q$. The $\blacklozenge$ 
 data points are the same as in (a).}
\label{linksL4}
\end{center}
\end{figure}

\begin{figure}[!t]
 \begin{center}
\includegraphics*[width=70mm]{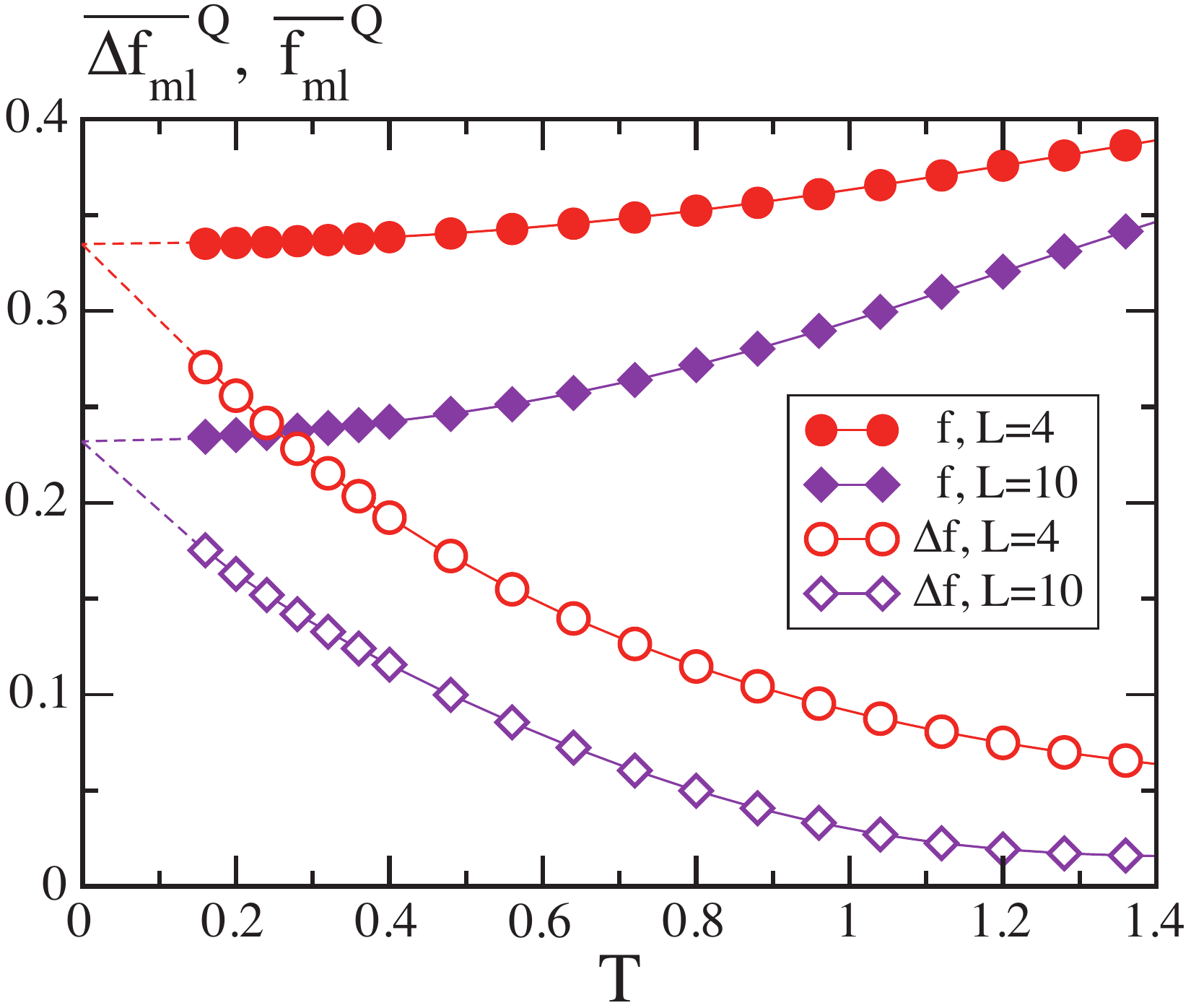}
\caption{(Color on line) Plots of  $\overline{\Delta f_{ml}}^Q$ and $\overline{ f_{ml}}^Q$ vs $T$ for $Q=1/4$ and the values of $L$ shown. The dotted lines are for our $T\rightarrow 0$ extrapolations.}
\label{bT}
\end{center}
\end{figure}

Plots of $L^{0.39} { f_{ml}}(q)$ vs $q$ are shown in Fig. \ref{links} for  EA systems of various sizes  at $T=0.16$.
However, departures from such scaling behavior can be observed, even over this limited range of system sizes,
in analogous plots (not shown) for $T$ as small as $T\gtrsim 0.3$. This effect is more clearly exhibited  in Fig. \ref{linksLm}(a), where
plots of  $\overline {f_{ml}}^Q$   vs $L$  are shown for $Q=1/2$ and various temperatures. 

A qualitatively different picture can be observed in Fig. \ref{linksLm}(b). In it,
plots of $\overline {\Delta f_{ml}}^Q$   vs $L$  are shown for the same $Q$ and $T$ as in Fig. \ref{linksLm}(a).  
All data points shown for $\overline {\Delta f_{ml}}^Q $  in Fig. \ref{linksLm}(b), which include temperatures up to $T\leq 0.8$, fall on straight lines. Consequently, $T\rightarrow 0$ extrapolations of their slope values is straightforward. From such extrapolations, we obtain $d-d_s=0.41(3)$.
 Within errors, this is in agreement with the value found in Refs. \onlinecite{kpy,py,jk} by a different method.

Incidentally, we note in Figs. \ref{linksLm}(a) and \ref{linksLm}(b) that whereas $\overline {f_{ml}}^Q$ decreases, $\overline{\Delta f_{ml}}^Q$ increases as $T$ decreases. This is as expected, because whereas the number of excitations (and thus $\overline {f_{ml}}^Q$) decreases as the temperature decreases, the \emph{cost} ($\overline{\Delta f_{ml}}^Q$) of creating an excitation  increases as the temperature decreases (since higher temperatures imply a larger number of  mismatched links to  start with).

For $Q=1/4$, plots of  $\overline { f_{ml}}^Q$ and $\overline {\Delta f_{ml}}^Q$   vs $L$  and various temperatures are shown in Figs. \ref{linksL4}(a) and \ref{linksL4}(b), respectively. Proceeding as above (for $Q=1/2$), we arrive at $d-d_s=0.41(3)$. We thus infer this number to hold independently of $Q$ in the $Q\lesssim 1/2$ interval.

 We can alternatively do a $T\rightarrow 0$ extrapolation  of $\overline {\Delta f_{ml}}^Q$ and $\overline {f_{ml}}^Q$ for each value of $L$, as shown in Fig. \ref{bT} for $Q=1/4$ and $L=4$ and $10$. Extrapolations from both curves meet, within errors, at the same point, as expected. The points thus obtained for $Q=1/2$ [$Q=1/4$] are plotted in Figs. \ref{linksLm}(a) and \ref{linksLm}(b) [Figs. \ref{linksL4}(a) and \ref{linksL4}(b)]. 
From  Log-log plots of the zero temperature curves thus obtained, we also obtained $d-d_s=0.41(3)$.

\section{conclusions}
\label{conclusions}

We have reported data for $p(q)$ from averages over large sets (numbers are shown in Table I) of EA and SK systems at very low temperature. 
The data for $\overline {p}^Q$  improves our confidence level in the conclusion that $p(q\sim 0)$  is nonzero and system-size independent  in the EA model.  
Future generation of more accurate data that would give $p(0)\sim 1/L^\theta$ and $\theta > 0.04$ in the $3\leq L\leq 10$ range of $L$ values is rather unlikely.
Thus, the conclusion KPY had reached,\cite{kpy} that the EA model exhibits a clear trend away from the droplet scenario, is strengthened. Furthermore, our results are consistent with $p(0)/T \rightarrow  0.233(4)$ as $T\rightarrow 0$ in the EA model (and $p(0)/T \rightarrow  0.51(3)$ as $T\rightarrow 0$ in the SK model).

We have studied the fraction of link mismatches, $\overline{\Delta f_{ml}}^Q$, 
 it \emph{costs}  to create an excitation 
with a $-Q<q<Q$ value. For a wide range of temperatures in the spin-glass phase, $\overline{\Delta f_{ml}}^Q$
seems to vanish, as in the TNT picture,\cite{halfway} in the macroscopic limit.
Data points for $\overline{\Delta f_{ml}}^Q$ (for $Q=1/4$ and $1/2$) are consistent with 
$\overline{\Delta f_{ml}}^Q \sim 1/L^{d-d_s}$ for all $T\lesssim T_{sg}$.  Furthermore, in agreement with results obtained in Refs. \onlinecite{kpy,py,jk} by a different method, we  find  
$d-d_s \rightarrow 0.41(3)$ as $T\rightarrow 0$. 

\acknowledgments
We thank the Centro de Supercomputaci\'on y Bioinform\'atica and  Laboratorio de M\'etodos Num\'ericos, both at Universidad de M\'alaga, for much computer time.
Funding from the Ministerio de Econom\'ia y Competitividad of Spain, through Grant FIS2009-08451,  is gratefully acknowledged.

{
\appendix

\section{Error bars}
\label{confidence}

 We show here how 
statistical errors $\Delta (N_s)$ for $\overline{p}^Q$ depend on $N_s$, on system size, on $Q$ and on $T$ for $T\ll 1$.

Consider first the rms deviation $\delta p(q)$ of the probability density $ p_{\cal J}(q) $ from its
average over different samples. It is plotted  
vs $\mid q \mid$ in Fig. 3 of Ref. [\onlinecite{us}] for  EA systems of various sizes at $T=0.1$.
Because there is no self-averaging, $\delta p \gg p$, except near $q=1$.
In addition, $\delta p(q)/p(q)$ does not decrease as $L$ increases. This has an unwanted implication, namely, 
fractional statistical errors in $p(q)$ do not decrease
as system size increases if $N_s$ remains constant.

 To start, let $F_{\cal J}(q_1,q_2)\equiv   p_{\cal J}(q_1)p_{\cal J}(q_2)$, 
\begin{equation}
G_{\cal J}(q\mid Q)  \equiv \int_{0}^{Q} dq_1\int_{0}^{Q} dq_2\; \delta (q_2-q_1-q)F_{\cal J}(q_1,q_2),
\label{defg}
\end{equation}
and let $G(q\mid Q)$ be the average of $G_{\cal J}(q\mid Q) $ over samples.
We can then write,
\begin{equation}
\Delta (1)^2=Q^{-2}\int_{-Q}^Qdq\;G(q\mid Q)-(\overline{p}^Q)^2
\end{equation}
follows.
Now, let 
\begin{equation}
\int_{-Q}^Qdq\;G(q\mid Q) \equiv wG(0\mid Q)
\end{equation}
define $w$.
We also note, 
\begin{equation}
G(0\mid Q)=  Q[(\overline{p}^Q)^2+\overline{(\delta p)^2}^Q].
\label{G0}
\end{equation}

Here, the first term is much smaller than the second one for $T\ll 0.5$ and all $Q\lesssim 1/2$, in both the EA and SK models.
This comes from the fact that, whereas $\overline{p}^Q$ vanishes as $T\rightarrow 0$,  $\overline{(\delta p)^2}^Q$ does not. Therefore,
$\Delta^2 (1)\simeq (w/Q)\overline{(\delta p)^2}^Q$ for $T\ll 0.5$ and all $Q\lesssim 1/2$, whence,
\begin{equation}
\Delta (N_s) \simeq  \overline{\delta p}^Q \sqrt {\frac{w}{QN_s}},
\label{err}
\end{equation}
follows immediately. 
This is clearly consistent with non-self-averaging. It shows that $\Delta $ is, at least for the values of $L$ we study here, independent of $L$, for both the EA and SK models. Equation (\ref{err}) also shows how much  precision  is gained by averaging $p(q)$ over $-Q<q<Q$.  

\begin{figure}[!t]
 \begin{center}
\includegraphics*[width=70mm]{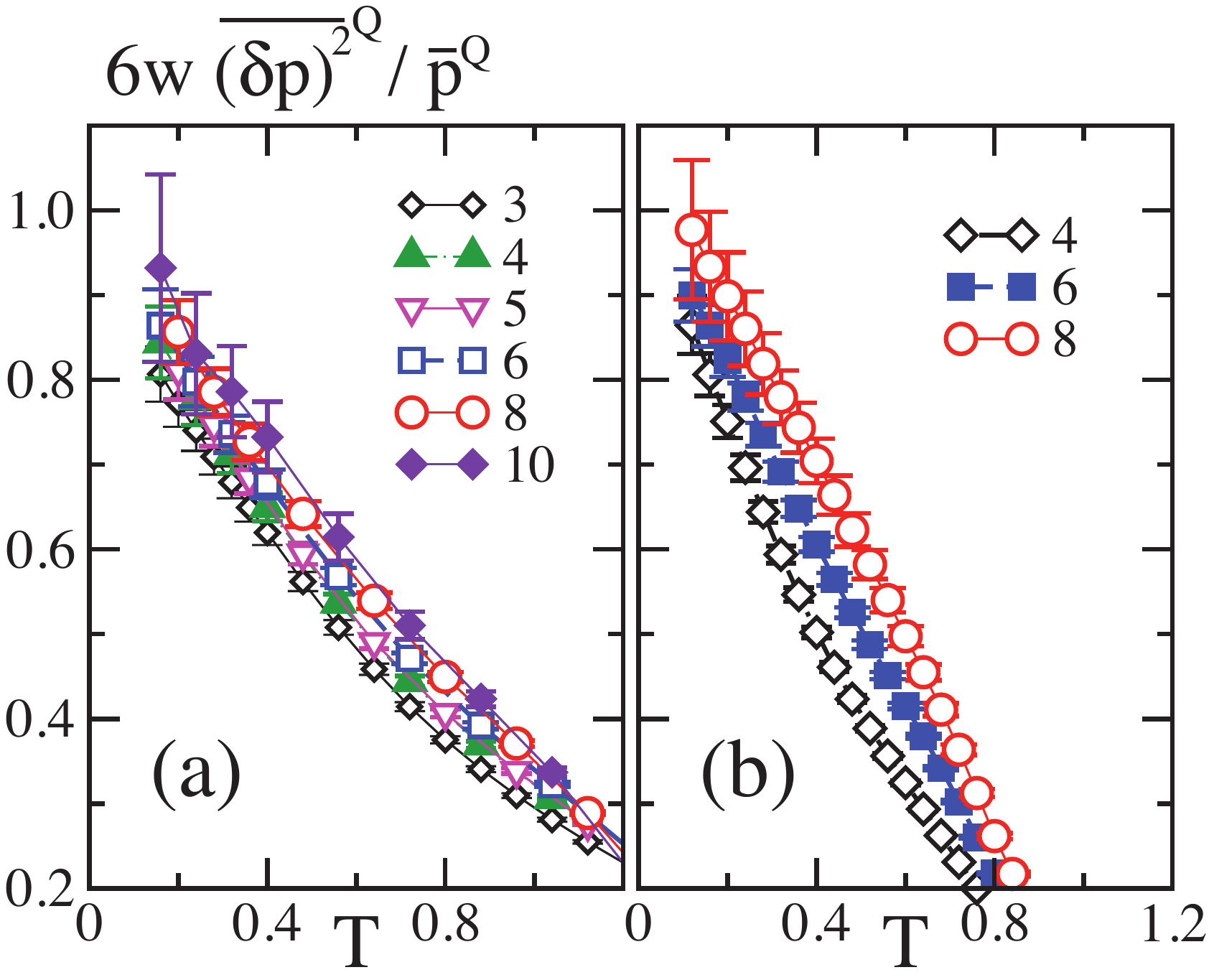}
\caption{(Color online)(a) Plots of $6w\overline{(\delta p)^2 }^Q/\overline{p}^Q$ vs $T$ for EA systems of $L$ linear sizes, as shown, and $Q=1/4$. (b) Same as in (a) but for the SK model. }
\label{alpha}
\end{center}
\end{figure}

We can substitute into Eq. (\ref{err}) the low temperature values of $w\overline{(\delta p)^2}^Q/\overline{ p}^Q$  from Figs. \ref{alpha}(a)  and \ref{alpha}(b) for the EA and SK models, respectively. Further substitutions of $\overline{p}^Q$ from Sec. \ref{averagep} give
\begin{equation}
\Delta (N_s) \simeq c\sqrt{\frac{T} {QN_s}},  
\label{DDD}
\end{equation}
where $c\simeq 0.2, 0.3$  for the EA and SK models, respectively, at low temperatures. This is the desired expression.

\begin{figure}[!]
 \begin{center}
\includegraphics*[width=70mm]{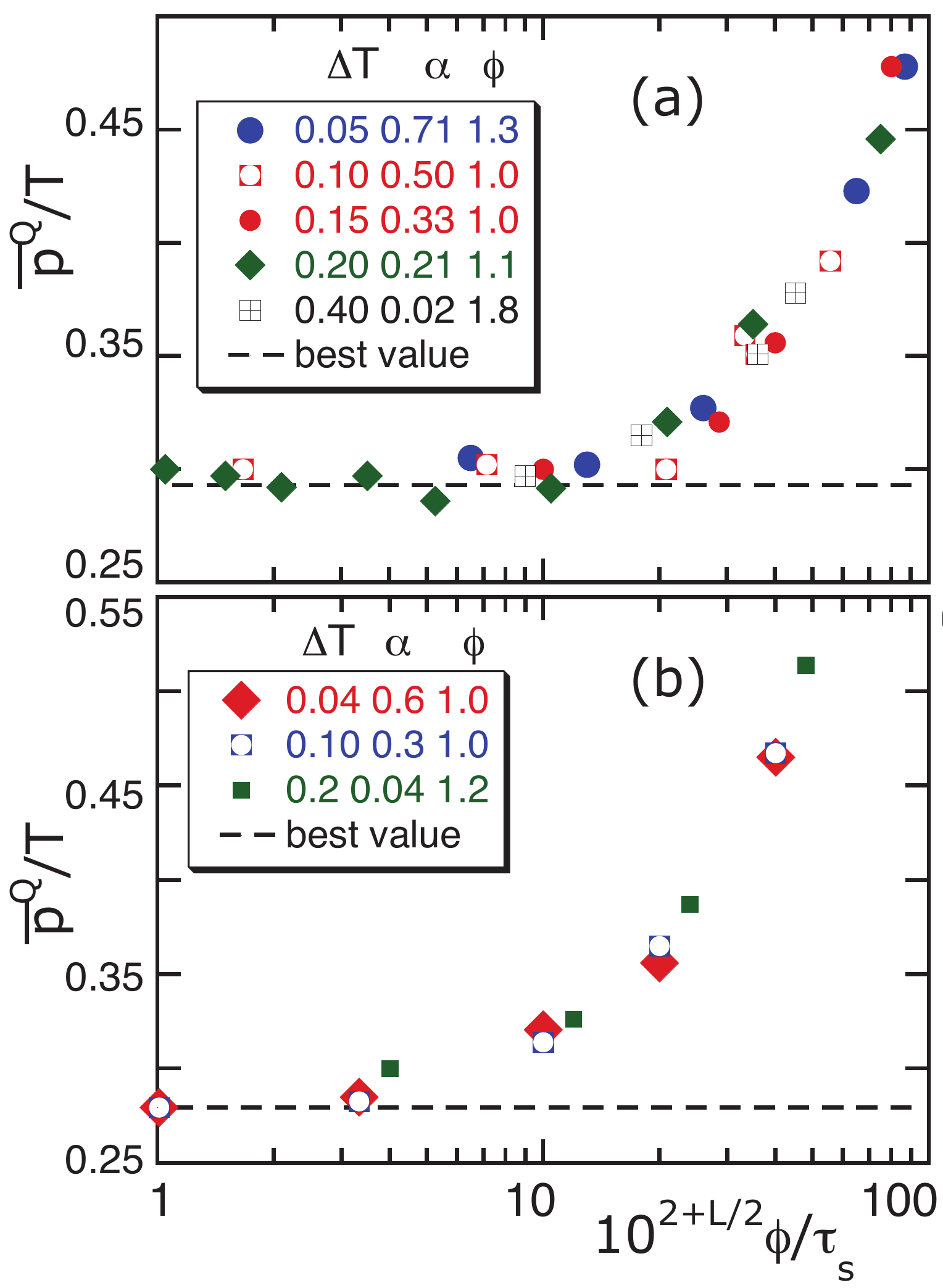}
\caption{(Color online) (a) Plots of the probability $\overline{p}^Q/T$ vs  $10^{2+L/2}\phi /\tau_s$ for $Q=1/2$ and at least $10^4$ sample EA systems of $6^3$ spins at $T=0.4$.
$\phi$ is a scaling number which is chosen so all points fall as closely as possible to a single curve. 
The values of $\Delta T$, swap success rate $\alpha$ for swaps between systems at $T$ and $T+\Delta T$,  and $\phi$ are as shown.
The coldest system in the tempered MC set up from which these data points follow was at $T=0.4$. (b) Same as in (a) but for systems of $8^3$ spins at $T=0.2$, which were the coldest ones in the parallel tempered MC set up from which these data points follow.}
\label{alfa1}
\end{center}
\end{figure}

\section{ Swap success rate and equilibration}
\label{swaprates}

We show here (i) how we choose the success rate, $\alpha$, for state swapping  (that is, for exchanging spin configurations) between two systems, and (ii) how we checked equilibration was achieved in our simulations.

We first derive an expression for $\alpha$. In the parallel tempered MC algorithm,\cite{tmc,tmc3} the probability, $P(ss \mid \Delta E )$, for state swapping to take place between  systems $1$ and $2$, at temperatures $T_1$, $T_2$,  where $\Delta E=E_2-E_1$, is given by,
$P(ss \mid \Delta E )=1$  if $ \Delta E\leq 0 $, but
$P(ss \mid \Delta E )=\exp (-\Delta \beta \Delta E)$ if $ \Delta E > 0$.
Now, 
\begin{equation}
\alpha =\int d\Delta E\;P(ss\mid \Delta E) \;P(\Delta E).
\label{alphaeq} 
\end{equation}
In thermal equilibrium, the probability that the energy of systems at $T_1$ and $T_2$ differ by $\Delta E$ is given by,
\begin{equation}
P( \Delta E)  \propto  \int dx\;e^{-x^2/2\sigma^2}e^{-(x-y)^2/2\sigma^2},
\end{equation}
where, neglecting variations in the specific heat (per spin) $c$ in the $T_1<T<T_2$ range,
$N\sigma^2$ is the mean square energy deviation coming from thermal fluctuations at both $T_1$ and $T_2$, 
\begin{equation}
x=\frac{E-\langle E \rangle}{\sqrt{N}} \quad\mathrm {and}\quad y=-\frac{\Delta E}{\sqrt{N}}+c\sqrt{N}\Delta T.
\end{equation}
It then follows that,
\begin{equation}
P( \Delta E)=(2\sqrt{\pi N} \sigma)^{-1}e^{-y^2/4\sigma^2}.
\end{equation}
Substitution into Eq. (\ref{alphaeq}) yields, assuming $\Delta T\ll T_1<T_2$,
\begin{equation}
\alpha \simeq   \operatorname{erfc} (\Delta T\sqrt {cN}/2T), \\
\label{ocho}
\end{equation}
[$ \operatorname{erfc}(\gamma )=(2/\sqrt{\pi}\int_{\gamma}^\infty dx\;\exp(-x^2)$].

A choice of $\alpha\approx 0.5$ might seem to lead to efficient MC simulations, which, using  Eq. (\ref{ocho}), would lead $\Delta T\sqrt {cN}/2T\approx 0.48$.
Note however that increasing $\Delta T$ does make $\alpha$ smaller, but it also implies fewer 
random steps need be taken by a given state in order to travel from a system at the minimum  temperature to one at the maximum temperature. 
Furthermore, smaller temperature differences imply fewer systems to be simulated, which leads to further computer time saving. 

This point is illustrated in Fig. \ref{alfa1}(a), where plots of $\overline{p}^Q/T$ vs  $10^{5}\phi /\tau_s$ are shown for
EA systems of $6\times 6\times 6$ spins at $T=0.4$. These data points come from tempered MC runs of sets of \emph{equally spaced} temperatures. Values of the  swap success rate, $\alpha$, between the two systems at the lowest pair of temperatures are given for each $\Delta T$, given by $T_{n+1}-T_n$, in Fig. \ref{alfa1}(a). 
From the values of $\phi$ given in Fig. \ref{alfa1}(a), we conclude that values of $\alpha$ as small as $0.2$ do not lead significantly slower simulations. Figure Fig. \ref{alfa1}(b) is as \ref{alfa1}(a) but for $L=8$, $T_{min}=0.1$, and $x$ axis values are for $10^6/\tau_s$. Note even an $\alpha$ value as small as $0.04$ only slows simulations down by $20\%$.
}


\begin{thebibliography}{99}

\bibitem{us}J. F. Fern\'andez and J. J. Alonso, Phys. Rev. B \textbf{86}, 140402(R) (2012).

\bibitem{EA}S. F. Edwards and P. W. Anderson, J. Phys. F \textbf{5}, 965 (1975).

\bibitem{SK}D. Sherrington and S. Kirkpatrick, Phys. Rev. Lett.
\textbf{32}, 1792 (1975).

\bibitem{seedD}See Eq. (7) in, M. M\'ezard, G. Parisi, N. Sourlas, G. Toulouse, and M. Virasoro, Phys. Rev. Lett. \textbf{52}, 1156 (1984). 


\bibitem{libro}M. M\'ezard, G. Parisi, and M. Virasoro, \emph{Spin Glass Theory
and Beyond} (World Scientific, Singapore, 2004). 



\bibitem{yu}The number of spikes beyond some threshold height in large sets of system samples
was found to increase in the SK model but remain approximately constant in the
EA model as $L$ increases, in B. Yucesoy, H. G. Katzgraber, and J. Machta, Phys. Rev. Lett. \textbf{109}, 177204 (2012).

\bibitem{paris}A. Billoire, L. A. Fern\'andez, A. Maiorano, E. Marinari, V. Mart\'{i}n-Mayor, G. Parisi, F. Ricci-Tersenghi,
J. J. Ruiz-Lorenzo, and D. Yllanes, arXiv:con-mat/1211.0843 (2012).  Note, however, that low temperature behavior is inferred here from high temperature data, where critical effects persist in fairly large EA systems in 3D.

\bibitem{bokil} M. A. Moore, H. Bokil, and B. Drossel, Phys. Rev. Lett. \textbf{81}, 4252 (1998).


\bibitem{kpy}H. G. Katzgraber, M. Palassini, and A. P. Young,
Phys. Rev. B, \textbf{63}, 184422 (2001) (referred to as KPY).

\bibitem{droplet}W. L. McMillan, J. Phys. C \textbf{17}, 3179 (1984).
D. S. Fisher and D. A. Huse, Phys. Rev. B \textbf {38}, 386 (1988); M. A. Moore and A. J. Bray, J. Phys. C \textbf{18}, L699 (1985);
M. A. Moore, J. Phys A \textbf{38}, L783 (2006); see also, A. A. Middleton, Phys. Rev. B \textbf{63}, 060202 (2001).

\bibitem{middleton} A note of caution on how deceptive finite-size behavior may sometimes be is given by
A. A. Middleton, in arXiv:cond-mat/1303.2253 (2013), where 
strong finite size corrections in a toy droplet model are examined.

\bibitem{halfway}F. Krz\c{a}kala and O. C. Martin, Phys. Rev. Lett. \textbf{85}, 3013 (2000).

\bibitem{py}M. Palassini and A. P. Young,  Phys. Rev. Lett. \textbf{85}, 3017 (2000) (referred to as PY).

\bibitem{jk}T. J\"org and H. G. Katzgraber, Phys. Rev. Lett. \textbf{ 101}, 197205 (2008). 


 
\bibitem{opposite}P. Contucci, C. Giardin\`a, C. Giberti, G. Parisi, and C. Vernia, Phys. Rev. Lett. \textbf {99}, 057206 (2007); G. Parisi and F. Ricci-Tersenghi, Philos. Mag. \textbf{92}, 341 (2012).

\bibitem{massive}Surface areas of large scale excitations in rather large systems (up to $L=32$)
were also examined by
R. Alvarez Ba\~nos, A. Cruz, L. A. Fern\'andez, J. M. Gil-Narvion, A. Gordillo-Guerrero, M Guidetti, A. Maiorano, F. Mantovani, E. Marinari, V. Mart\'{\i}n-Mayor, J. Monforte-Garc\'{\i}a, A. Mu\~noz Sudupe, D. Navarro, G. Parisi, S. P\'erez-Gaviro, J. J. Ruiz-Lorenzo, S F Schifano, B. Seoane, A. Taranc\'on, R. Tripiccione, and D. Yllanes, J. Stat. Mech. \textbf{2010}, 06026 (2010). These simulations were carried out not too far below the critical temperature, and moving into the low temperature regime by increasing the size of EA systems in 3D at constant temperature is expected to require huge system sizes   
for the EA model in 3D. Not surprisingly, it was concluded that the results obtained were ``pre-asymptotic'' (that is, far from the macroscopic regime).


\bibitem{TcEA}H. G. Katzgraber, M. K\"orner, and A. P. Young, Phys. Rev. B \textbf{73}, 224432 (2006).

\bibitem{tmc}K. Hukushima and K. Nemoto, J. Phys. Soc. Jpn. \textbf{65}, 1604 (1996); see also J. F. Fern\'andez, Phys. Rev. B \textbf{82}, 144436, (2010); J. J. Alonso and J. F. Fern\'andez, Phys. Rev. B \textbf{81}, 064408 (2010).

\bibitem{tmc3}For a critique of the  parallel tempered MC method, see, J. Machta, Phys. Rev. E \textbf{80}, 056706 (2009). 

\bibitem{vani}J. Vannimenus, G. Toulouse, and G. Parisi, J. Phys. (Paris) \textbf{42}, 565 (1981).


\bibitem{Q}The slight downward displacement of KPY's data points with respect to our own can be accounted for by the
slight difference between their $Q=0.20$ value and our own $Q=1/4$.

\bibitem{toulo}G. Toulouse, Comm. Phys. \textbf{2}, 115 (1977).



\end{thebibliography}
\end{document}